\documentclass[12pt]{article}
\usepackage[T1]{fontenc}
\usepackage[latin9]{inputenc}
\usepackage{float}
\usepackage{amsmath}
\usepackage{graphicx}
\usepackage{amssymb}

\makeatletter

\@ifundefined{definecolor}
 {\usepackage{color}}{}

\usepackage{graphics}
\usepackage{epsfig}
\usepackage{latexsym}
\usepackage{latexsym}

\newcommand{\be}{\begin{eqnarray}}
\newcommand{\ee}{\end{eqnarray}}
\newcommand{\bra}[1]{\mbox{$\langle\, #1 \mid$}}

\newcommand{\ket}[1]{\mbox{$\mid #1\,\rangle$}}

\newcommand{\pro}[2]{\mbox{$\langle\, #1 \mid #2\,\rangle$}}

\renewcommand{\d}{\mbox{{\rm d}}}

\newcommand{\vect}[1]{\mbox{\boldmath$#1$}}

\newcommand{\midvect}[1]{\mbox{\boldmath${\textstyle #1}$}}
\def\gsim{\lower0.5ex\hbox{$\:\buildrel >\over\sim\:$}}
\def\lsim{\lower0.5ex\hbox{$\:\buildrel <\over\sim\:$}}

\makeatother

\begin{document}

% Title of the paper
\author{Khireddine Nouicer\thanks{khnouicer@univ-jijel.dz } \\
%EndAName
\textit{Laboratory of Theoretical Physics and Department of Physics,}\\
\textit{\ Faculty of Sciences, University of Jijel}\\
\textit{\ Bp 98, Ouled Aissa, 18000 Jijel, Algeria.}\\
Youssef Sabri\thanks{sabri@fias.uni-frankfurt.de}\\
\textit{Frankfurt Institute for Advanced Studies, }\\
\textit{Johann Wolfgang Goethe University,}\\
\textit{Ruth-Moufang-Str. 1 60438 Frankfurt am Main, Germany.}}

\title{{\Large The Casimir force in noncommutative Randall-Sundrum models}}

\date{\today}
\maketitle
\begin{abstract}
In this paper we study the effect of spacetime noncommutativity in
the 5-dimensional Randall-Sundrum brane worlds on the Casimir force
acting on a pair of parallel plates. We show that the presence of
a noncommutative scale length affects the nature of the Casimir force
for small plate separation. Using accurate experimental bounds for
the Casimir force in parallel plate geometry, we find an upper bound
for the noncommutative cutoff of the order of $10^{3}$ TeV, and that
the size of the interbrane distance in RSI model is approximately
given by $kR\lesssim20.5$ and $kR\lesssim18.4$ for $k=10^{19}$
GeV and $k=10^{16}$GeV, respectively.

11.25.Wx, 11.25.Mj, 11.10.Kk, 11.10.Nx{\small \par}

Casimir Force, Warped Extra Dimensions, Randall-Sundrum Models{\small \par}

\end{abstract}
%\vskip -2.5cm
%%%%%%%%%%%%%%%%%%%%%%%%%%%%%%%%%%%%%%%%%%%%%%%%%%%%%%%%%%%%%%%%%%%%

\section{Introduction}

\label{sec:intro} %%%%%%%%%%%%%%%%%%%%%%%%%%%%%%%%%%%%%%%%%%%%%%%%%%%%%%%%%%%%%%%%%%%%

String theory suggests that the spacetime we live in might be higher
dimensional and theories with extra dimensions have received much
attention in high energy physics, especially in the context of hierarchy
problems and cosmology \cite{Horava-witten,Brax}. In this scenario,
it is very likely that our four-dimensional universe is a hypersurface
called brane embedded in a higher dimensional spacetime called a bulk.
According to this brane world scenario, gravity and other exotic matter
can propagate in the bulk, while all matter and gauge interactions
are confined on the brane. On the other hand, one of the most important
questions in particle physics today is how to explain the huge discrepancy
between the gravity scale $M_{{\rm Pl}}$ and the electro-weak scale
$M_{{\rm EW}}$.

The first braneworld scenario addressing the problem of the mass hierarchy
is the Arkani-Hamed-Dimopoulos-Dvali (ADD) model with $n$ flat compactified
large extra dimensions of size $L$ \cite{Arkani-Hamed}. In this
scenario, the observed Planck scale is the product of the fundamental
Planck scale of the full volume of the bulk and the volume of the
extra dimensions \begin{equation}
M_{{\rm Pl}}^{2}=L^{n}M_{4+n}^{2+n}\,.\end{equation}
Therefore, in the ADD model the fundamental Planck scale could be
much smaller than the Planck scale and could be comparable to the
electroweak scale if more than one extra dimension are used.

Major progress has been made by Randall and Sundrum (RS) who considered
a warped 5-dimensional bulk geometry compactified on a $S^{1}/Z_{2}$
manifold, and provide an alternative approach to explain the huge
discrepancy between the electroweak scale and the Planck scale \cite{Randall}.
The geometry is that of a 5-dimensional Anti-de-Sitter space ($AdS_{5}$),
which is a space of constant negative curvature. There are two RS
scenarios: in the RSI scenario, the geometry contains two 4-D flat
branes localized at $y=0$ and $y=L$, respectively. The Planck brane
(hidden) with positive tension where gravity is localized, and the
TeV brane (visible) with negative tension where all standard model
fields are confined, that bound an extra dimension which is compactified
to $S^{1}/Z_{2}$ orbifold. In this scenario, the 5-D gravity is sourced
by the two 4-D flat branes with opposite tensions and a finely tuned
non-vanishing 5D cosmological constant. Since the two branes are completely
flat, the induced metric at every point along the extra dimension
has to be the ordinary flat 4-D Minkowski metric, and the components
of the 5-D metric depend only on the fifth coordinate. Thus, the most
general spacetime metric satisfying these properties is given by \begin{equation}
ds^{2}=e^{-2k\mid y\mid}\eta_{\mu\nu}dx_{\mu}dx_{\nu}-dy^{2},\label{metric}\end{equation}
where $y$ is the physical distance along the extra dimension. The
quantity $e^{-2k|y|}$ is the warp factor which measures the degree
of curvature along the extra dimension. The parameter $k$, which
governs the degree of curvature of the $AdS_{5}$ space, is assumed
to be of the order of the Planck scale. Considering the fluctuations
of the metric $(\ref{metric})$, one obtains the expression for the
effective 4-D gravity scale as a function of the RS parameters \begin{equation}
M_{{\rm Pl}}^{2}=\frac{M_{5}^{3}}{k}\left(1-e^{-2\pi kR}\right),\label{rel-MPL}\end{equation}
where $M_{5}$ is the fundamental 5-D gravity scale. 

In the second Randall-Sundrum model (RSII), the bulk is infinite and
there is only one brane with positive tension located at $y=0$. In
this scenario there is no mechanism to solve the hierarchy problem.
The spectrum of RSII is continuous and consist of all $m>0$ KK modes,
and there are no ${\cal O}$(TeV) signatures for this model at the
colliders. The infinite extra dimension makes a finite contribution
to the 5-D volume because of the warp factor, and the effective size
of the extra dimension is $1/k$.

On the other hand, it is established that string theory induced noncommutative
(NC) geometry \cite{chu,Witten,Seiberg} provides an effective minimal
length theory to study short distance physics. Also one of the most
interesting consequences of all promising quantum gravity candidates
is the existence of a minimal observable length on the order of the
Planck length. The idea of a minimal length can be modeled in terms
of a quantized spacetime and goes back to the early days of quantum
field theory \cite{snyder} (see also \cite{connes}-\cite{bondia}).
However, in a scenario with extra dimensions, it is expected that
the effects of the string inspired NC scale become important at the
same energy scale at which the effective extra dimensional models
predict new physics. Thus, the NC minimal length should be lowered
down to TeV energies \cite{Hossenfelder-1}.

A powerful tool to probe the existence and physical implication of
extra dimensions, by studying the quantum fluctuations of vacuum in
higher dimensional spacetime, is the Casimir effect \cite{Casimir}.
The later is a fundamental aspect of quantum field theory in confined
geometries and depends crucially on the dimensionality and topology
of the spacetime. Actually the precision of the measurement of the
Casimir effect has been greatly improved \cite{Lamoreaux}, making
the Casimir effect to be remarkably observable and a trustworthy manifestation
of the quantum fluctuations. The Casimir effect as the physical manifestation
of zero-point energy has received great attention, and has been extensively
studied. The effect has been investigated in the context of string
theory \cite{Fabinger}, and with its relation in stabilizing the
radion in the Randall-Sundrum model \cite{Garriga}. The Casimir effect
for parallel plates geometry has been also studied for a scalar field
in spacetime with compactified universal extra dimensions \cite{Hossenfelder}
and for massless bulk field in the Randall-Sundrum models \cite{Turan},
and for massive and massless scalar field in a Randall-Sundrum type
braneworld consisting of a single brane extended by one compact extra
dimension \cite{Linares}. Recently, the standard Casimir effect in
the presence of a minimal length, arising from NC spacetimes and generalized
uncertainty principle (GUP), has been also considered \cite{Debabi,Nouicer,Casadio,Panella}.

In this paper we investigate the possible effects that a NC scale
length might have on the Casimir force for massless bulk scalar fields
confined between two parallel plates in the 5-dimensional RS brane
worlds. In our setup, in order to maintain Lorentz invariance, the
noncommutativity between coordinates is only implemented on the brane.
For simplicity, we adopt the scalar field-photon analogy to calculate
the Casimir force due to the electromagnetic field \cite{Milton}.
In section II, we give a brief review of the realization of spacetime
noncommutativity in the RS models. In section III, we calculate explicitly
to leading order, in the case where the interbrane distance is larger
than the NC scale length, the renormalized Casimir energy and force
acting on the plates. In section IV, we perform a numerical analysis
of our analytical expressions, and discuss the consequences of the
presence of the NC cutoff on the nature of the Casimir force. Our
concluding remarks and summarized in section 5.

\section{Noncommutative spacetimes}

It is a common lore that any consistent formulation of quantum gravity
leads to the appearance of a minimal length of the order of the Planck
length. A particular implementation of such a minimal length can be
realized in the so called NC field theory on spacetimes where the
coordinates satisfy the following structure relations \begin{equation}
\left[x^{\mu},x^{\nu}\right]=i\Theta^{\mu\nu},\label{structure}\end{equation}
and where $\Theta^{\mu\nu}$ is an antisymmetric Lorentz tensor. An
immediate consequence of these relations is the following generalized
uncertainty principle (GUP) and minimal length \begin{equation}
\varDelta x^{\mu}\varDelta x^{\nu}\geq\rvert\Theta^{\mu\nu}\rvert,\qquad\varDelta x\sim\sqrt{\rvert\Theta\rvert}.\end{equation}
These relations go back to the early days of quantum field theory
and appeared for the first time in the pioneering work of Snyder \cite{snyder}.
These are also the structure relations satisfied, in the context of
string theory, by the endpoints of open strings in non trivial RR
flux backgrounds \cite{chu,Witten,Seiberg}.

In this section, we implement the spacetime noncommutativity on the
brane in the RS scenario using the method based on the coherent states
representation \cite{Smailagic}. 

A $D$-dimensional NC brane can be defined in terms of spacetime coordinates
$x^{\mu}$ (where $\mu=1,2,\ldots,D$) on the brane which satisfy
the structure relations $(\ref{structure})$ and \begin{equation}
\left[x^{\mu},y\right]=0,\end{equation}
where $y$ is the extra dimension.

A suitable rotation transforms the tensor $\Theta^{\mu\nu}$ into
a block-diagonal form, \begin{equation}
\hat{\Theta}^{\mu\nu}={\rm diag}\left(\hat{\Theta}_{1},\hat{\Theta}_{2},\ldots,\hat{\Theta}_{D/2}\right)\ ,\end{equation}
 where \begin{equation}
\hat{\Theta}_{i}=\,\left({\begin{array}{cc}
0 & \theta_{i}\\[2ex]
-\theta_{i} & 0\end{array}}\right)\ .\label{tens}\end{equation}
 In order to have full noncommutativity, one needs to work in a spacetime
that has an even number of dimensions. Then, the $D=2\, d$ coordinates
can be represented by $d$ two-vectors: \begin{equation}
\hat{x}^{\mu}=\left(\vec{\hat{x}}_{1},\vec{\hat{x}}_{2},\ldots,\vec{\hat{x}}_{d}\right),\end{equation}
where $\vec{\hat{x}}_{i}\equiv\left(\hat{z}_{1i},\hat{z}_{2i}\right)$
are two-vectors in the $i$-th NC plane such that \begin{equation}
\left[\hat{z}_{1i},\hat{z}_{2i}\right]=i\theta_{i}.\label{commutation}\end{equation}
Let us then construct a set of commuting ladder operators from the
NC spacetime coordinates. Indeed, the ladder operators for the $i$-th
plane are defined by \begin{equation}
\begin{array}{l}
\hat{a}_{i}=\frac{1}{\sqrt{2}}\left(\hat{z}_{1i}+i\,\hat{z}_{2i}\right),\\
\\\hat{a}_{i}^{\dagger}=\frac{1}{\sqrt{2}}\left(\hat{z}_{1i}-i\,\hat{z}_{2i}\right)\ ,\end{array}\end{equation}
which satisfy the canonical commutation rules \begin{equation}
\left[\hat{a}_{i},\hat{a}_{j}^{\dagger}\right]=\delta_{ij}\,\theta_{i}\ .\end{equation}
The operators $\hat{a}_{i}$ and $\hat{a}_{i}^{\dagger}$ can be established
as annihilation and creation operators of a harmonic oscillator, and
coherent states form a suitable basis in the Fock space. Normalized
coherent states can now be defined as \begin{equation}
\ket{\alpha}=\prod_{i}\,\exp\left[\frac{1}{\theta_{i}}\,\left({\alpha}_{i}^{*}\,\hat{a}_{i}-\alpha_{i}\,\hat{a}_{i}^{\dagger}\right)\right]\ket{0}\ ,\pro{\alpha}{\alpha}=1,\label{coherent states}\end{equation}
where $\ket{0}$ is the vacuum state annihilated by all $\hat{a}_{i}$. 

Using expectations values over the coherent states, we can associate
ordinary functions to any operator-valued function as follows \begin{equation}
F(\alpha)=\langle\alpha\mid F(\hat{\vect{x}})\mid\alpha\rangle,\end{equation}
and by this way, it results a generalized plane wave given by \begin{equation}
\bra{\alpha}\exp\left[i\,\sum_{i=1}^{d}\,\left(\vec{p}\cdot\vec{\hat{x}}\right)_{i}\right]\ket{\alpha}=\exp\left\{ -\sum_{i=1}^{d}\left[\frac{1}{4}\theta_{i}\vect{p}_{i}^{2}+i\left(\vec{p}\cdot\vec{{x}}\right)_{i}\right]\right\} ,\label{plane-wave}\end{equation}
where $\vect{p}_{i}$ is the momentum canonically conjugate to the
spacetime coordinate $x_{i}$. In a 4-D NC brane we have two such
NC planes ($d=2$) and it is worth noting that, for complete spacetime
noncommutativity and $\theta_{i}>0$, the damping factors are positive
both for time and space momentum components, regardless if a Minkowski
or Euclidean metric is used. 

In the following, for simplicity, we will assume that the parameters
which describe noncommutativity in the $i$-th plane in Eq.~(\ref{commutation})
are all the same, $\theta_{i}=\theta>0$. Using the generalized plane
waves defined in Eq.$(\ref{plane-wave})$, we express now a scalar
field operator in a NC 4-D Minkowski spacetime as \begin{equation}
\Phi(t,\vec{x})=\int{\d^{3}p}\left[\hat{b}_{\vec{p}}u_{{\rm nc}}(t,\vec{x})+\hat{b}_{\vec{p}}^{\dag}u_{{\rm nc}}^{*}(t,\vec{x})\right],\label{exp}\end{equation}
where the modes $u_{{\rm nc}}(t,\vec{x})$ are given by \begin{equation}
u_{{\rm nc}}(t,\vec{x})=\frac{e^{-\frac{\theta}{4}\left(\omega^{2}+p^{2}\right)}e^{i\left(\vec{p}\cdot\vec{x}-\omega t\right)}}{(2\,\pi)^{3/2}\sqrt{2\omega}},\end{equation}
and the ladder operators $\hat{b}_{\vec{p}}$ and $\hat{b}_{\vec{p}}^{\dag}$
obey the canonical commutation relation \begin{equation}
\left[\hat{b}_{\vec{p}},\hat{b}_{{\vec{p}}'}^{\dag}\right]=\delta^{(3)}(\vec{p}-\vec{p}').\end{equation}
Let us now consider the energy operator of the massless scalar field
\begin{equation}
\hat{H}=\frac{1}{2}\int\d^{3}x\left[\Pi^{2}+\left(\vec{\nabla}\Phi\right)^{2}\right],\end{equation}
where $\Pi(t,\vect{x})=\dot{\Phi}(t,\vect{x})$ is the conjugate momentum
field. Using the expansion of the scalar field given by Eq.$(\ref{exp})$,
we obtain \begin{equation}
\hat{H}=\int\frac{\d^{3}p}{(2\,\pi)^{3}}\,\omega(p)\, e^{-{\theta}\,\omega^{2}}\left(\hat{b}_{{\vec{p}}}^{\dag}\,\hat{b}_{\vec{p}}+\frac{1}{2}\right).\end{equation}
The zero-point energy defined by $E_{0}=\langle0|H|0\rangle$, is
then \begin{equation}
E_{0}=\int_{0}^{\infty}\frac{p^{2}\,\d p}{(2\,\pi)^{2}}e^{-\theta\omega^{2}}\omega(p),\end{equation}
which is finite due to the Gaussian damping factor which dominates
at large $p$.

After these brief review of canonical NC scalar field in the coherent
state representation, we turn now to study the Casimir effect in NC
RSI and RSII Brane scenarios.

\section{The Casimir effect}

In this section, we evaluate the Casimir force for a massless scalar
field confined between two parallel plates in the NC 5-dimensional
RSI braneworld scenario. We impose the standard Dirichlet boundary
condition on the wave vector in the direction orthogonal to the plates
$\left.\Psi\left(x^{\mu},y\right)\right|_{\partial\Omega}=0$, where
$\partial\Omega$ is the location of the plates on the brane. The
KK spectrum for the bulk scalar field has been discussed in the literature
\cite{Rattazzi}, and for the non-zero modes it consists of a tower
of modes exponentially suppressed and given by \begin{equation}
k_{N}\thickapprox\kappa(N+\frac{1}{4}),\label{extra-mode}\end{equation}
where $\kappa=\pi ke^{-\pi kR}$. The approximation in the above equation
is valid asymptotically for $N\gg1$, or equivalently $\pi kR\gg1,$
but is already very accurate even for $N=1$, where the deviation
from the actual value is about 0.03. 

The Casimir energy density per unit plate area will be then obtained
by summing up the zero-point energy, where the frequency of the vacuum
fluctuations is \begin{equation}
\omega_{nN}=c\sqrt{\vect{k}_{\perp}^{2}+\left(\frac{\pi n}{L}\right)^{2}+k_{N}^{2}}\,,\end{equation}
where $k_{N}$ is defined in Eq.~(\ref{extra-mode}), $\vect{k}_{\perp}^{2}=\sqrt{k_{x}^{2}+k_{y}^{2}}$
, $L$ is the distance between the plates and the integers $n$ label
the normal modes between the plates. Therefore, the total energy inside
the plates reads \begin{equation}
{\cal E}_{{\rm C,p}}^{{\rm RSI}}=A\hbar\int\frac{d^{2}\midvect{k}_{\perp}}{(2\pi)^{2}}\left(pc\left\{ \sum_{n,N=0}^{\infty}\right\} ^{*}\omega_{nN}e^{-\theta\hbar^{2}\omega_{nN}^{2}}-pc\sum_{N=1}^{\infty}\omega_{0N}e^{-\theta\hbar^{2}\omega_{0N}^{2}}\right),\label{frequency-sum}\end{equation}
where $A$ is the area of one plate, and we have subtracted the contribution
of the modes polarized in the direction of the brane \cite{Ambjorn}.
The factor $p$ accounts for the possible polarizations of the photon.
In our case $p=3$ in 4-D spacetime, and we have taken into account
the overall factor of 2 for the volume of the orbifold, which cancel
the usual $1/2$ factor. The first term in Eq.$(\ref{frequency-sum})$
contains the NC contribution without the RSI scenario $(k_{N}=0)$,
and the star indicates that the term with $N=0$ is excluded. The
first term in the Casimir energy can be decomposed as \begin{eqnarray}
p\hbar c\left\{ \sum_{n,N=0}^{\infty}\right\} ^{*}\omega_{nN}e^{-\theta\hbar^{2}\omega_{nN}^{2}} & = & p'\hbar c\sum_{n=1}^{\infty}\sqrt{\vect{k}_{\perp}^{2}+\left(\frac{\pi n}{L}\right)^{2}}e^{-\hbar^{2}\theta\left(\vect{k}_{\perp}^{2}+\left(\frac{\pi n}{L}\right)^{2}\right)}+p\hbar c\sum_{n,N=0}^{\infty}\omega_{nN}e^{-\theta\hbar^{2}\omega_{nN}^{2}}\nonumber \\
 & - & p\hbar c\sum_{n=0}^{\infty}\omega_{n0}e^{-\theta\hbar^{2}\omega_{n0}^{2}},\end{eqnarray}
where the term with the polarization factor $p'$, corresponds to
the case of a massless scalar field localized on the NC spacetime
without the RS scenario.

Let us start our calculation by considering integrals of the following
form \begin{equation}
I=\int\frac{d^{2}\vect{k}_{\perp}}{\left(2\pi\right)^{2}}e^{-l_{nc}^{2}\left(\vect{k}_{\perp}^{2}+r^{2}\right)}\sqrt{\left(\vect{k}_{\perp}^{2}+r^{2}\right)},\label{eq:int1}\end{equation}
where we have defined the fundamental NC scale $l_{nc}=\hbar\sqrt{\theta}$.
Using polar coordinates, the integral (\ref{eq:int1}) becomes \begin{eqnarray}
I & = & \frac{e^{-l_{nc}^{2}r^{2}}}{4\pi}\int_{r^{2}}^{\infty}dye^{-l_{{\rm nc}}^{2}y}\left(y+r^{2}\right)^{1/2}\nonumber \\
 & = & \frac{l_{{\rm nc}}^{-3}}{4\pi}\Gamma\left(\frac{3}{2},l_{{\rm nc}}^{2}r^{2}\right),\label{eq:int2}\end{eqnarray}
where $\Gamma(\alpha,z)$ is the incomplete gamma function. Using
( \ref{eq:int2}) and the following formula \begin{equation}
\Gamma(\alpha,x)=\Gamma\left(\alpha\right)-\sum_{s=0}^{\infty}\frac{(-1)^{s}x^{\alpha+s}}{s!(\alpha+s)},\end{equation}
the different contributions in Eq.$(\ref{frequency-sum})$, which
we denote respectively by $I_{j}$ ($j=1.2,3,4$) are now given by
\begin{equation}
I_{1}=\frac{l_{{\rm nc}}^{-3}}{4\pi}\sum_{n=0}^{\infty}\Gamma\left(3/2\right)-\frac{l_{{\rm nc}}^{-3}}{4\pi}\sum_{s=0}^{\infty}\left(\frac{l_{{\rm nc}}^{2}\pi^{2}}{L^{2}}\right)^{3/2+s}\frac{(-1)^{s}}{s!(3/2+s)}\zeta(-3-2s),\end{equation}
 \begin{equation}
I_{2}=\frac{l_{{\rm nc}}^{-3}}{4\pi}\sum_{n=0,N=0}^{\infty}\Gamma\left(3/2\right)-\frac{1}{4\pi}\sum_{s=0}^{\infty}\left({l_{{\rm nc}}}\right)^{2s}\frac{(-1)^{s}}{s!(3/2+s)}E_{2}\left(-\frac{3}{2}-s;\frac{\pi^{2}}{L^{2}},\kappa^{2};0,\frac{1}{4}\right),\end{equation}
 \begin{equation}
I_{3}=\frac{l_{{\rm nc}}^{-3}}{4\pi}\sum_{n=0}^{\infty}\Gamma\left(3/2\right)-\frac{1}{4\pi}\sum_{s=0}^{\infty}\frac{(-1)^{s}}{s!(3/2+s)}E_{1}^{\kappa^{2}/16}\left(-\frac{3}{2}-s;\frac{\pi^{2}}{L^{2}};0\right),\end{equation}
 \begin{equation}
I_{4}=\frac{l_{{\rm nc}}^{-3}}{4\pi}\sum_{N=1}^{\infty}\Gamma\left(3/2\right)-\frac{\kappa^{3}}{4\pi}\sum_{s=0}^{\infty}\frac{(-1)^{s}\kappa^{2s}}{s!(3/2+s)}\left(\zeta_{H}(-3-2s,\frac{1}{4})-(1/4)^{2s+3}\right),\end{equation}
where \begin{eqnarray}
E_{2}\left(s;a_{1},a_{2};c_{1},c_{2}\right) & = & \sum_{n_{1},n_{2}=0}^{\infty}\left[a_{1}(n_{1}+c_{1})^{2}+a_{2}(n_{2}+c_{2})\right]^{-s},\\
E_{1}^{c}\left(s;a_{1};c_{1}\right) & = & \sum_{n_{=}0}^{\infty}\left[a_{1}(n_{1}+c_{1})^{2}+c^{2}\right]^{-s},\end{eqnarray}
are zeta functions of the Epstein-Hurwitz type and $\zeta_{H}(s,q)=\sum_{n=0}^{\infty}(n+q)^{-s}$
is the Hurwitz zeta function. Substituting in Eq.$(\ref{frequency-sum})$
we obtain \begin{eqnarray}
{\cal E}_{{\rm C,p}}^{{\rm RSI}} & = & -\frac{{A}\hbar}{4\pi}\sum_{s=0}^{\infty}\frac{(-1)^{s}l_{{\rm nc}}^{2s}}{s!(3/2+s)}\Bigg[p'c\left(\pi/L\right)^{2s+3}\zeta(-2s-3)+pcE_{2}\left(-\frac{3}{2}-s;\frac{\pi^{2}}{L^{2}},\kappa^{2};0,\frac{1}{4}\right)\nonumber \\
 & - & pc\kappa^{2s+3}\left(\zeta_{H}(-2s-3,\frac{1}{4})-(1/4)^{2s+3}\right)-pcE_{1}^{\kappa^{2}/16}\left(-\frac{3}{2}-s;\frac{\pi^{2}}{L^{2}};0\right)\Bigg],\label{eq:EnRS}\end{eqnarray}
where the first term with the polarization factor $p'$, is the expression
of the Casimir energy on NC spacetime without the RSI brane scenario.

Let us then calculate the total Casimir energy without the plates,
given by\begin{flalign}
{\cal E}_{{\rm C,np}}^{{\rm RSI}} & ={A}\hbar pc\int\frac{d^{3}\vect{k}}{(2\pi)^{3}}\Bigg[\sum_{N=0}^{\infty}e^{-l_{{\rm nc}}^{2}(\vect{k}^{2}+\kappa^{2}(N+1/4)^{2})}\nonumber \\
\times & \sqrt{\vect{k}^{2}+\kappa^{2}(N+1/4)^{2}}-e^{-l_{{\rm nc}}^{2}(\vect{k}^{2}+\kappa^{2}/16)}\sqrt{\vect{k}^{2}+\kappa^{2}/16}.\label{eq:eout}\end{flalign}
The integrals over $k$ can be calculated directly in terms of the
modified Bessel functions, but in order to easily identify the divergent
terms of the same nature as the ones in Eq.(\ref{eq:EnRS}), we follow
another method. Let us start with the second term in Eq.$(\ref{eq:eout})$
and denote it by $J$. We first decompose the integral over the three
dimensional momentum to a product of integrals along the longitudinal
and transverse momenta, and then applying the Schwinger proper-time
representation to obtain \begin{equation}
J=\frac{1}{8\pi^{2}\Gamma(-1/2)}\int_{0}^{\infty}dss^{-3/2}e^{-\frac{\kappa^{2}}{16}(l_{{\rm nc}}^{2}+s)}\int_{-\infty}^{+\infty}{dk_{z}}e^{-(l_{{\rm nc}}^{2}+s)k_{z}^{2}}\int_{0}^{\infty}dye^{-(l_{{\rm nc}}^{2}+s)y}.\end{equation}
Performing the Gaussian integrals over the momenta and finally integrating
over $s$ we obtain \begin{equation}
J=\frac{2\kappa^{4}e^{-l_{{\rm nc}}^{2}\kappa^{2}/16}}{16^{3}\pi^{3/2}}\sum_{k=0}^{\infty}\left(-\frac{l_{{\rm nc}}^{2}\kappa^{2}}{16}\right)^{k}\frac{\Gamma(-k-2)}{k!\Gamma(-k-1/2)}.\label{fterm}\end{equation}
The first term in Eq.$(\ref{eq:eout})$ is obtained just by making
the substitution $\kappa^{2}/16\longrightarrow\kappa^{2}(N+1/4)^{2}$
in Eq.$(\ref{fterm})$, and performing the summation over $N$. Collecting
the resulting expressions we obtain the Casimir energy in the RSI
scenario without the plates \begin{equation}
{\cal E}_{{\rm C,np}}^{{\rm RSI}}=A\frac{\hbar pc\kappa^{4}}{8\pi^{3/2}}\sum_{k=0}^{\infty}\left(-l_{{\rm nc}}\kappa\right)^{2k}\frac{\Gamma(-k-2)}{k!\Gamma(-k-1/2)}\left[\zeta_{H}\left(-4-2k,\frac{1}{4}\right)-\frac{(-1)^{k}}{(16)^{k+2}}\right].\end{equation}
Here we note the appearance of a natural perturbation parameter, $l_{{\rm nc}}\kappa,$
which is in fact the ratio between the NC length and the interbrane
distance in RSI model. In the following, we consider the natural choice,
$l_{{\rm nc}}\kappa<1,$ which means that the NC length is smaller
than the distance between the two branes. Then, to leading order in
$l_{{\rm nc}}\kappa$, the Casimir energy without the plates becomes
\begin{equation}
{\cal E}_{{\rm C,np}}^{{\rm RSI}}=\frac{A\hbar pc\kappa^{4}}{8\pi^{3/2}}\frac{\Gamma(-2)}{\Gamma(-1/2)}\left[\zeta_{H}(-4,1/4)-\frac{1}{256}\right]+\frac{A\hbar pcl_{{\rm nc}}^{2}\kappa^{6}}{20\pi^{3/2}}\frac{\Gamma(-3)}{\Gamma(-3/2)}\left[\zeta_{H}(-6,1/4)+\frac{1}{4096}\right],\label{energy-out}\end{equation}
while the Casimir energy inside the plates follows from (\ref{eq:EnRS}),
\begin{eqnarray}
{\cal E}_{{\rm C,p}}^{{\rm RSI}} & = & -\frac{{A}\hbar}{6\pi}\Bigg[p'c\left(\frac{\pi}{L}\right)^{3}\zeta(-3)+pcE_{2}\left(-\frac{3}{2};\frac{\pi^{2}}{L^{2}},\kappa^{2};0,\frac{1}{4}\right)\nonumber \\
 & - & \kappa^{3}\left(\zeta_{H}\left(-3,\frac{1}{4}\right)-\frac{1}{64}\right)-pcE_{1}^{\kappa^{2}/16}\left(-\frac{3}{2};\frac{\pi^{2}}{L^{2}},0\right)\Bigg]\nonumber \\
 & + & \frac{A\hbar l_{{\rm nc}}^{2}}{10\pi}\Bigg[p'c\left(\frac{\pi}{L}\right)^{5}\zeta(-5)+pcE_{2}\left(-\frac{5}{2};\frac{\pi^{2}}{L^{2}},\kappa^{2};0,\frac{1}{4}\right)\nonumber \\
 & - & \kappa^{5}\left(\zeta_{H}\left(-5,\frac{1}{4}\right)-\frac{1}{1024}\right)-pcE_{1}^{\kappa^{2}/16}\left(-\frac{5}{2};\frac{\pi^{2}}{L^{2}},0\right)\Bigg].\label{eq:EC}\end{eqnarray}

\noindent Now, we use the following expansion of the Epstein-Hurwitz
zeta function type \cite{Elizalde92,Elizalde94} \begin{eqnarray}
E_{2}(s;a_{1},a_{2};c_{1},c_{2}) & = & \frac{a_{2}^{-s}}{\Gamma(s)}\sum_{m=0}^{\infty}\frac{(-1)^{m}\Gamma(s+m)}{m!}\left(\frac{a_{1}}{a_{2}}\right)^{m}\zeta_{H}(-2m,c_{1})\nonumber \\
 &  & \times\zeta_{H}(2s+2m,c_{2})+\frac{a_{2}^{1/2-s}}{2}\sqrt{\frac{\pi}{a_{1}}}\frac{\Gamma(s-1/2)}{\Gamma(s)}\zeta_{H}(2s-1,c_{2})\nonumber \\
 &  & +\frac{2\pi^{s}}{\Gamma(s)}\cos(2\pi c_{1})a_{1}^{-s/2-1/4}a_{2}^{-s/2+1/4}\sum_{n_{1}=1}^{\infty}\sum_{n_{2}=0}^{\infty}n_{1}^{s-1/2}\label{E2}\\
 &  & \times(n_{2}+c_{2})^{-s+1/2}K_{s-1/2}\left(2\pi\sqrt{\frac{a_{2}}{a_{1}}}n_{1}(n_{2}+c_{2})\right)\;,\nonumber \end{eqnarray}
 and \begin{eqnarray}
E_{1}^{c}(s;a_{1};c_{1}) & = & \frac{c^{-s}}{\Gamma(s)}\sum_{m=0}^{\infty}\frac{(-1)^{m}\Gamma(s+m)}{m!}\left(\frac{a_{1}}{c}\right)^{m}\zeta_{H}(-2m,c_{1})+\frac{c^{1/2-s}}{2}\sqrt{\frac{\pi}{a_{1}}}\frac{\Gamma(s-1/2)}{\Gamma(s)}\label{eq:}\\
 &  & +\frac{2\pi^{s}}{\Gamma(s)}a_{1}^{-s/2-1/4}c^{-s/2+1/4}\sum_{n_{1}=1}^{\infty}\cos(2\pi n_{1}c_{1})n_{1}^{s-1/2}K_{s-1/2}\left(2\pi\sqrt{\frac{c}{a_{1}}}n\right),\nonumber \end{eqnarray}
where $K_{\nu}\left(x\right)$ is the modified Bessel function of
the second kind, and $\zeta_{H}(s,q)=\sum_{n=0}^{\infty}(n+q)^{-s}$
is the Hurwitz zeta function.

\noindent Substituting these relations in Eq.$(\ref{eq:EC})$ with
$s=-3/2$ and $s=-5/2$, respectively, it is easy to observe that
the divergent terms, which are proportional to $\Gamma(-2)$ and $\Gamma(-3)$,
are exactly the same ones in $(\ref{energy-out})$. Then, the renormalized
Casimir energy takes the explicit form\begin{flalign}
{\cal {E}}_{{\rm C}}^{RSI} & ={\cal E}_{{\rm C,p}}^{{\rm RSI}}-{\cal E}_{{\rm C,np}}^{{\rm RSI}}L\nonumber \\
 & =-\frac{A\hbar p'c\pi^{2}}{6L^{3}}\left[\zeta\left(-3\right)-\frac{3\pi^{2}}{5}\frac{l_{{\rm nc}}^{2}}{L^{2}}\zeta\left(-5\right)\right]\nonumber \\
 & -\frac{A\hbar pc\kappa^{2}}{6\pi}\Bigg[\frac{p\kappa}{128}-\frac{p+2}{2}\kappa\zeta_{H}\left(-3,\frac{1}{4}\right)-\frac{\kappa}{256}-\frac{3p}{32L\pi}\sum_{n=1}\frac{K_{2}\left(L\kappa n/2\right)}{n^{2}}\nonumber \\
 & +\frac{3p}{2L\pi}\sum_{n=1}\sum_{N=0}\frac{(N+1/4)^{2}}{n^{2}}K_{2}\left(2L\kappa n\left(N+\frac{1}{4}\right)\right)\Bigg]\nonumber \\
 & +\frac{A\hbar cl_{{\rm nc}}^{2}\kappa^{3}}{10\pi}\Bigg[\frac{p\kappa^{3}}{2048}-\frac{p+2}{2}\kappa^{2}\zeta_{H}\left(-5.\frac{1}{4}\right)+\frac{\kappa^{3}}{1024}-\frac{15p}{256L^{2}\pi}\sum_{n=1}\frac{K_{3}\left(L\kappa n/2\right)}{n^{3}}\nonumber \\
 & +\frac{15p}{4L^{2}\pi}\sum_{n=1}\sum_{N=0}\frac{(N+1/4)^{3}}{n^{3}}K_{3}\left(2L\kappa n\left(N+\frac{1}{4}\right)\right)\Bigg].\label{eq:Ecasimir}\end{flalign}
Let us now, compute the Casimir force per unit plate area defined
by $F_{{\rm C}}^{RSI}=-\frac{\partial({\cal {E}}_{{\rm C}}^{RSI}/A)}{\partial L}$.
Defining the reduced Casimir force $\mathcal{F}_{{\rm C}}^{RSI}=F_{{\rm C}}^{RSI}/\kappa^{4}$
and the dimensionless parameters $\lambda=l_{{\rm nc}}\kappa$ and
$\mu=L\kappa$, we finally obtain \begin{eqnarray}
\mathcal{F}_{{\rm C}}^{RSI} & = & \mathcal{F}_{{\rm C}}^{noRS}-\frac{\hbar pc}{4\pi^{2}}\frac{1}{\mu^{2}}\left\lbrace \sum_{n=1}^{\infty}\sum_{N=0}^{\infty}\frac{(N+1/4)^{2}}{n^{2}}K_{2}(2L\kappa n(N+1/4))\right.\nonumber \\
 & + & \mu\sum_{n=1}^{\infty}\sum_{N=0}^{\infty}\frac{(N+1/4)^{3}}{n}\left[K_{1}\left(2\mu n\left(N+1/4\right)\right)+K_{3}\left(2\mu n\left(N+1/4\right)\right)\right]\label{force}\\
 &  & \left.-\frac{1}{16}\sum_{n=1}^{\infty}\frac{K_{2}\left(\mu n/2\right)}{n^{2}}-\frac{\mu}{64}\sum_{n=1}^{\infty}\left(\frac{K_{1}\left(\mu n/2\right)+K_{3}\left(\mu n/2\right)}{n}\right)\right\rbrace \nonumber \\
 & + & \frac{3\hbar pc}{4\pi^{2}}\frac{\lambda^{2}}{\mu^{3}}\left[\sum_{n=1}\sum_{N=0}\frac{(N+1/4)^{2}}{n^{3}}K_{3}\left(2\mu n\left(N+1/4\right)\right)-\frac{1}{64}\sum_{n=1}\frac{K_{3}\left(\mu n/2\right)}{n^{3}}\right]\nonumber \\
 & + & \frac{3\hbar pc}{8\pi^{2}}\frac{\lambda^{2}}{\mu^{2}}\Bigg[\sum_{n=1}\sum_{N=0}\frac{(N+1/4)^{4}}{n^{2}}\left(K_{2}\left(2\mu n\left(N+1/4\right)\right)+K_{4}\left(2\mu n\left(N+1/4\right)\right)\right)\nonumber \\
 & - & \frac{1}{256}\sum_{n=1}^{\infty}\left(\frac{K_{2}\left(\mu n/2\right)+K_{4}\left(\mu n/2\right)}{n^{2}}\right)\Bigg],\nonumber \end{eqnarray}
where we have used the differentiation rule for the Bessel functions
$\partial_{z}K_{\nu}(z)=-\frac{1}{2}\left[K_{\nu-1}(z)+K_{\nu+1}(z)\right]$.
The first term $\mathcal{F}_{C}^{noRS}$ in Eq.(\ref{force}) is the
reduced NC Casimir force without the braneworld scenario, \begin{equation}
\mathcal{F}_{{\rm C}}^{noRS}=-\frac{\hbar p'c}{480}\frac{\pi^{2}}{\mu^{4}}-\frac{\hbar p'c\pi^{4}}{1008}\frac{\lambda^{2}}{\mu^{6}},\label{eq:NoRS}\end{equation}
which coincide with the expression already obtained in \cite{Debabi},
but differs from the one recently found in \cite{Casadio} using the
Euler-Maclaurin formula. Up to the order of perturbation we have used,
the NC contribution due to the RSI scenario given by the two last
terms in (\ref{force}) can be repulsive, and then stabilize the interbrane
distance. Let us note that the expression of the Casimir force given
by Eq.$(\ref{force})$ converges exponentially owing to the MacDonald's
representation of the Bessel function \begin{equation}
K_{\nu}(x)=\sqrt{\frac{\pi}{2x}}e^{-x}\sum_{n=0}^{\infty}\frac{1}{n!(8x)^{n}}\prod_{m=1}^{n}\left(4\nu^{2}-(2m-1)^{2}\right).\end{equation}
Then, it is enough for a numerical study of the Casimir force to consider
only the sum of the first terms.

\noindent Let us now derive an asymptotic expression for the Casimir
force valid for $\mu\gg1$. Since $K_{n}(x)\sim\sqrt{\pi/2x}\exp(-x)$
for large $x$, we may retain only the terms with $N=0$ and $n=1$
in Eq.(\ref{force}). Thus we obtain \begin{equation}
\mathcal{F}_{{\rm C}}^{RSI}=\mathcal{F}_{{\rm C}}^{noRS}-\frac{45\hbar pc}{512\pi^{3/2}}\frac{\lambda^{2}}{\mu^{5/2}}e^{-\mu/2}.\label{eq:asympt}\end{equation}
We observe, that in the commutative limit $(\lambda=0)$, there is
no RS contribution to the Casimir force, for $\mu\gg1$. Indeed, in
the regime $\mu\gg1$, the second term in Eq.(\ref{eq:asympt}) is
a purely NC effect. Taking into account that $\lambda$ plays the
role of a cutoff on the 3-dimensional brane, the correction term in
Eq.(\ref{eq:asympt}) is of the same form as the leading term in the
asymptotic expression of the Casimir force for a scalar field with
mass $m=\kappa/4$ confined between two parallel plates in 5-dimensional
spacetime \cite{Ambjorn}.

Now, we should proceed to the calculation of the Casimir force in
the 5-dimensional RSII model on NC spacetime coordinates. In the RSII
model the KK spectrum is continuous due to the suppression of the
second boundary at $y=\pi R$, and it consists of all $m>0$. Then,
the extra mode summation over $N$ in Eq.~(\ref{frequency-sum})
is turned into an integration with measure $dm/k$ \cite{Randall},
and the Casimir energy in the presence of the plates is \begin{equation}
{\cal E}_{{\rm C,p}}^{RSII}={\cal E}_{C}^{noRS}+\frac{A\hbar pc}{2}\int\frac{dm}{k}\sum_{n=1}^{\infty}\int\frac{d^{2}\midvect{k}_{\perp}}{(2\pi)^{2}}e^{-l_{{\rm nc}}^{2}\left(\vect{k}_{\perp}^{2}+\frac{\pi^{2}n^{2}}{a^{2}}+m^{2}\right)}\sqrt{\vect{k}_{\perp}^{2}+\frac{\pi^{2}n^{2}}{a^{2}}+m^{2}}\;.\end{equation}
Following the method of calculation used in the case of the 5-dimensional
RSI model, we obtain \begin{equation}
{\cal E}_{{\rm C,p}}^{RSII}={\cal E}_{C}^{noRS}+\frac{A\hbar pc}{8\pi kl_{{\rm nc}}^{3}}\int dm\sum_{n=1}^{\infty}\Gamma\left(\frac{3}{2},l_{{\rm nc}}^{2}\left(\frac{\pi^{2}n^{2}}{L^{2}}+m^{2}\right)\right).\end{equation}
Ignoring terms which do not contribute to the Casimir force and working
to leading order in the NC scale, we obtain \begin{equation}
{\cal E}_{{\rm C,p}}^{RSII}={\cal E}_{C}^{noRS}-\frac{A\hbar pc}{8\pi k}\int dm\left(\frac{2}{3}E_{1}^{m^{2}}\left(-\frac{3}{2};\frac{\pi^{2}}{L^{2}},0\right)-\frac{2l_{{\rm nc}}^{2}}{5}E_{1}^{m^{2}}\left(-\frac{5}{2};\frac{\pi^{2}}{L^{2}},0\right)\right).\end{equation}
Using Eq.(\ref{eq:}) and ignoring again terms independent on $L$,
we have \begin{eqnarray}
{\cal E}_{C,p}^{{\rm RSII}} & = & {\cal E}_{C}^{noRS}-\frac{A\hbar pc}{8\pi k}\int dm\Bigg(\frac{m^{4}L}{3\sqrt{\pi}}\frac{\Gamma(-2)}{\Gamma(-3/2)}+\frac{4m^{2}}{3L\sqrt{\pi}}\frac{1}{\Gamma(-3/2)}\sum_{n=1}^{\infty}\frac{K_{2}\left(2Lnm\right)}{n^{2}}\nonumber \\
 & - & \frac{m^{6}Ll_{{\rm nc}}^{2}}{5\sqrt{\pi}}\frac{\Gamma(-3)}{\Gamma(-5/2)}-\frac{4m^{3}l_{{\rm nc}}^{2}}{5L^{2}\sqrt{\pi}}\frac{1}{\Gamma(-5/2)}\sum_{n=1}\frac{K_{3}\left(2Lnm\right)}{n^{3}}\Bigg).\end{eqnarray}
This expression contains divergent terms and must be renormalized
by subtracting the contribution of the energy without the plates.
Indeed, the Casimir energy for the RSII model without the plates to
leading order in the NC scale takes the form \begin{equation}
{\cal E}_{{\rm C,np}}^{{\rm RSII}}=-\frac{A\hbar pc}{16\pi^{3/2}k}\int dmm^{4}\left(\frac{\Gamma(-2)}{\Gamma(-1/2)}-l_{{\rm nc}}^{2}m^{2}\left(\frac{\Gamma(-2)}{\Gamma(-1/2)}-\frac{\Gamma(-3)}{\Gamma(-3/2)}\right)\right).\end{equation}
Then, the renormalized Casimir energy reads as \begin{eqnarray}
{\cal E}_{C}^{RSII} & = & {\cal E}_{C,p}^{{\rm RSII}}-{\cal E}_{{\rm C,np}}^{{\rm RSII}}L\nonumber \\
 & = & {\cal E}_{C}^{{\rm noRS}}-\frac{A\hbar pc}{8\pi^{2}k}\sum_{n=1}^{\infty}\int dm\Bigg(\frac{m^{2}}{L}\frac{K_{2}\left(2Lnm\right)}{n^{2}}-\frac{3l_{{\rm nc}}^{2}m^{3}}{2L^{2}}\frac{K_{3}\left(2Lnm\right)}{n^{3}}\Bigg).\end{eqnarray}
Performing the integration over $m$ with the aid of the integral
$\int_{0}^{\infty}dmm^{q}K_{q}(\alpha m)=2^{q-1}\alpha^{-q-1}\sqrt{\pi}\Gamma\left(q+1/2\right)$,
we obtain\[
{\cal E}_{C}^{RSII}={\cal E}_{C}^{{\rm noRS}}-\frac{3\pi A\hbar pc}{128\pi^{2}kL^{4}}\left[\zeta\left(5\right)-\frac{15l_{{\rm nc}}^{2}}{4L}\zeta\left(7\right)\right].\]
Then, the corresponding Casimir force par unit area, for the RSII
model, takes the form \begin{equation}
F_{{\rm C}}^{RSII}=F_{{\rm C}}^{noRS}\left(1+\frac{45}{2}\frac{\hbar pc}{\pi^{3}kL}\left[\zeta(5)-\frac{15}{4}\left(\frac{l_{nc}}{L}\right)^{2}\zeta\left(7\right)\right]\right).\label{forceRSII}\end{equation}
Using the expression of $F_{{\rm C}}^{noRS}$ given by Eq.(\ref{eq:NoRS}),
we finally obtain to leading order in the NC scale length,

\begin{equation}
F_{{\rm C}}^{RSII}=F_{{\rm C}}^{Stand}\left\{ \left[1+\frac{45}{2}\frac{\hbar pc}{\pi^{3}kL}\zeta(5)\right]-\frac{75}{7}\frac{\hbar pc}{\pi kL}\left(\frac{l_{nc}}{L}\right)^{2}\left[\zeta\left(5\right)+\frac{63}{8\pi^{2}}\zeta\left(7\right)\right]\right\} ,\label{eq:RSII}\end{equation}
where $F_{C}^{Stand}=-\frac{\hbar p'c}{480}\frac{\pi^{2}}{L^{4}}$,
is the standard Casimir force. The second term in the first square
brackets is exactly the attractive contribution of the RSII scenario
without the minimal length \cite{Turan}, while the second square
brackets contain the NC repulsive RSII contribution, respectively.
However, by the GUP, we know that any physical characteristic length
of the model that is experimentally accessible, must be greater than
the minimal length. Then, we have $l_{nc}<L,$ and due to the factor
$1/kL,$ the correction term multiplying $F_{C}^{Stand}$ is of the
order of unity such that the Casimir force is always attractive.

\section{Numerical analysis}

In this section, we proceed to a numerical analysis of the Casimir
force in RSI brane scenario, particularly the implications of a non
zero NC scale length. In Fig.1, we show the variation of the reduced
Casimir force as a function of the dimensionless plate separation
$\mu$ for different values of the dimensionless NC scale length $\lambda$.
For the sake of comparison we use the experimental results $\mathcal{F}_{C}^{exp_{\pm}}=-K_{C}/\mu^{4}/\kappa^{4}$
where the Casimir factor is $K_{C}=\left(1.22\pm0.18\right)\cdot10^{-27}$
N/m$^{2}$ \cite{Lamoreaux}. We observe, that for $\lambda\lesssim10^{-2}$
all the curves coincide and lie within the experimental bounds, and
it is observed that there is no contribution coming from the RS scenario.
From Fig.1, we observe also the attractive character of the Casimir
force as it is expected from Eq.(\ref{eq:asympt}), valid for large
plate separation. Now, to scrutinize the nature of the Casimir force
for small plate separations, we have to extract some useful information
about the admissible values of the NC scale length. Using the experimental
bounds of the Casimir force given above, we solve $\mathcal{F}_{C}^{exp_{\pm}}=\mathcal{F}_{C}$
for the NC length $\lambda$. It should be noted that only with $\mathcal{F}_{C}^{exp_{-}}$
that one obtains real solutions in the large plate separation regime.
The resulting solution is plotted in Fig.2 as function of the plate
separation $\mu$. We observe the existence of two branches of solutions
with a gap between them. In fact, at the the first order approximation
used in our calculation, we obtain $\lambda=\sqrt{\frac{\mathcal{F}_{C}^{exp_{-}}-\mathcal{F}_{C}^{RSI}(\lambda=0)}{T}}$,
where $T$ is the sum of terms multiplying $\lambda$ in Eq.(\ref{force}),
and it is found that the denominator is negative for values $\mu$
lying in the gap. On the other hand, we have not used the experimental
bounds on the force, $\mathcal{F}_{C}^{exp_{+}}$or the arithmetic
average $\left(\mathcal{F}_{C}^{exp_{+}}+\mathcal{F}_{C}^{exp_{-}}\right)/2,$
since they reproduce only the values on the first branch. For our
purpose, we reject the first branch since it corresponds to very small
plate separations which are far from being probed experimentally,
and consider only the second branch starting at $\mu_{0}=3.0029.$
The later corresponds to a plate separation and a NC scale length,
of order $L_{0}\thickapprox10^{-19}$ m and $\left(l_{nc}\right)_{0}\thickapprox10^{-22}$
m, respectively. In conclusion, we can set the following bound on
the NC scale length, $\Lambda_{nc}\lesssim10^{3}$ TeV, where $\Lambda_{nc}=1/l_{nc}.$

\begin{figure}[H]
%\vskip -0.3in

\begin{centering}
\hspace*{-0.8cm} \includegraphics[width=12cm,height=8cm]{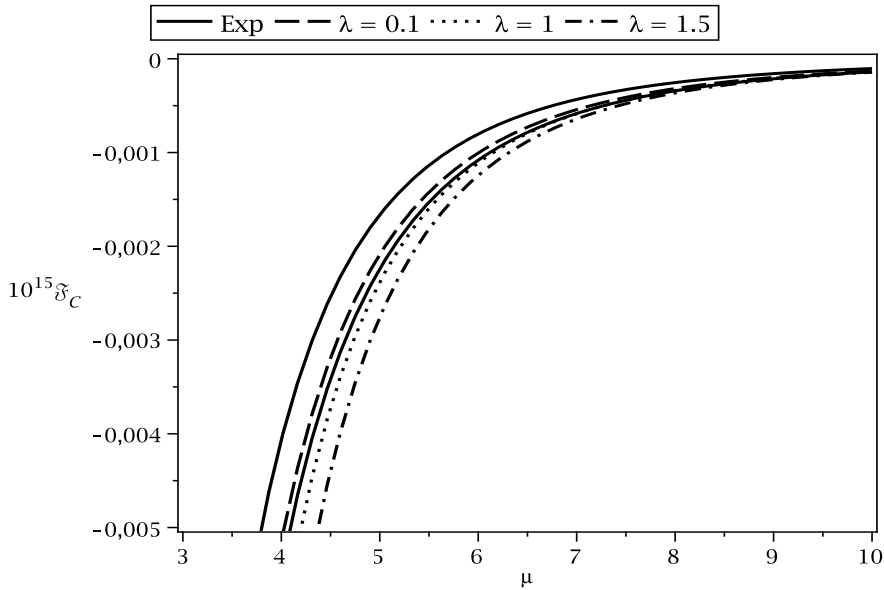}
\vskip -0.2in 
\par\end{centering}

\caption{The Casimir force versus large plate separation $\mu$ for different
values of the dimensionless NC scale $\lambda$. The black solid curves
are the experimental Casimir force $F_{C}^{exp_{\pm}}/\kappa^{4}$
with a $\pm15\%$ error level.}

\label{fig:Casimir-force} 
\end{figure}

\begin{figure}[htb]
%\vskip -0.3in

\begin{centering}
\hspace*{-0.8cm} \includegraphics[width=11cm,height=8cm]{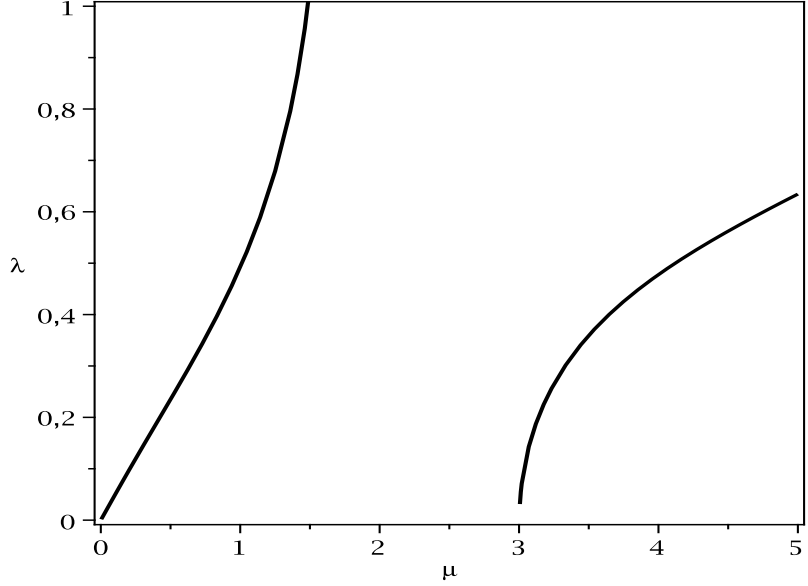}
\vskip -0.2in 
\par\end{centering}

\caption{The dimensionless NC length $\lambda$ as a function of the dimensionless
plate separation $\mu$ for $\mathcal{F}_{C}^{RSI}=\mathcal{F}_{C}^{exp_{-}}$. }

\label{fig:lambda} 
\end{figure}

The behavior of the Casimir force in the regime of small and intermediary
values of plate separation is shown in Fig.3. We observe that for
$l_{nc}\lesssim\left(l_{nc}\right)_{0}$, the Casimir force is attractive
in the whole region, and becomes repulsive for $l_{nc}>\left(l_{nc}\right)_{0}$.
Then, up to the order of perturbation we have used, the NC part of
the Casimir force contribute to the stabilization of the radion. However,
if we accept the experimental evidence that the Casimir force for
parallel plate geometry is always attractive, the NC cutoff is then
$\Lambda_{nc}\lesssim10^{3}$ TeV, which is stronger than the ones
available in the literature. Currently, the most robust lower bound
on the NC scale $\Lambda_{nc}\gtrsim1$$ $ TeV, comes from the $Z\longrightarrow\gamma\gamma$
decay originating from the renormalizable gauge sector of the noncommutative
standard model (NCSM) \cite{Buric}. However, a recent analysis using
the big-bang nucleosynthesis (BBN) restriction on the number of neutrinos
species gives $\Lambda_{nc}\gtrsim3$ TeV and $\Lambda_{nc}\gtrsim10^{3}$
TeV for $\Delta N_{\nu}=1$ and $\Delta N_{\nu}\lesssim0.2,$ respectively
\cite{Horvat}. 

\begin{figure}[H]
%\vskip -0.3in

\begin{centering}
\includegraphics[width=12cm,height=8cm]{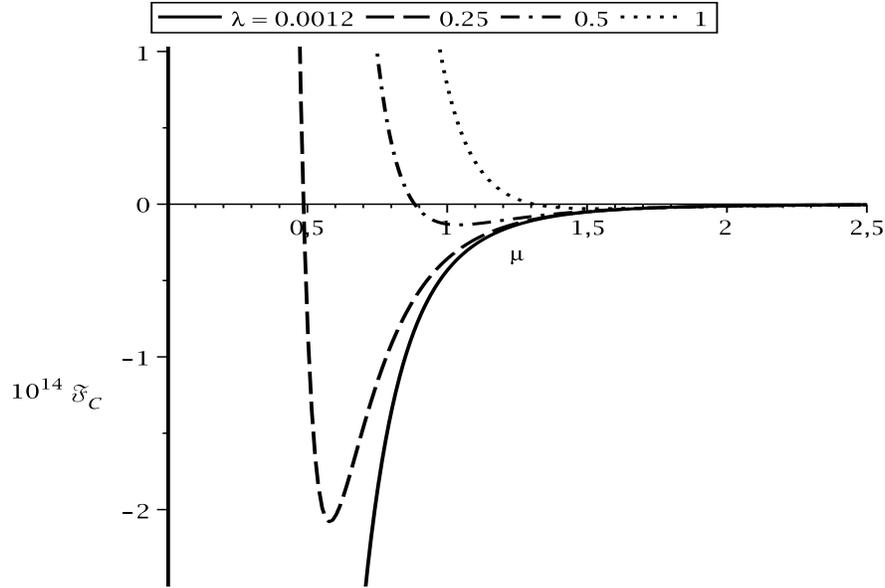}
\par\end{centering}

\vskip -0.2in 

\caption{The Casimir force as a function of the plate separation $\mu$ for
small NC length . }

\label{fig:parameter-kR} 
\end{figure}

Finally, we look for the possible values of the interbrane distance
$kR$ for non zero NC scale length. In Fig.4 we show the variation
of the Casimir force as function of $kR$ for $k=10^{19}$ GeV and
$k=10^{16}$ GeV, respectively. Our analysis is performed with the
plate separation $L=0.5$ $\mu m,$ and different values of the NC
scale length $l_{nc}$. In the left and right panel we observe that,
for a NC scale smaller than $10^{-2}$ $\mu$m, the upper bound for
$kR$ is around 20.5 and 18.4, respectively. 

\begin{figure}[H]
%\vskip -0.3in

\begin{centering}
\includegraphics[width=8cm,height=10cm]{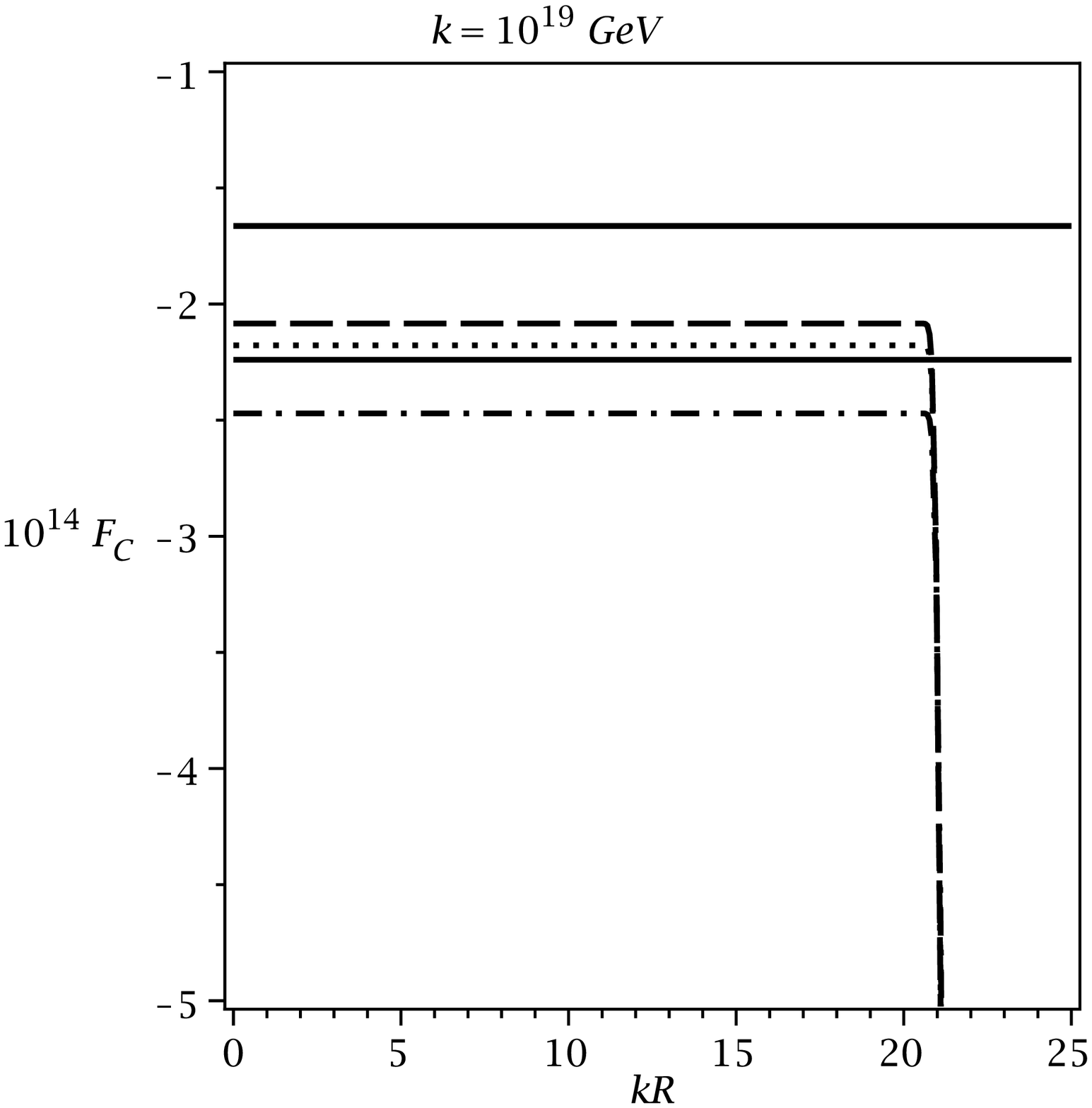}\includegraphics[width=8cm,height=10cm]{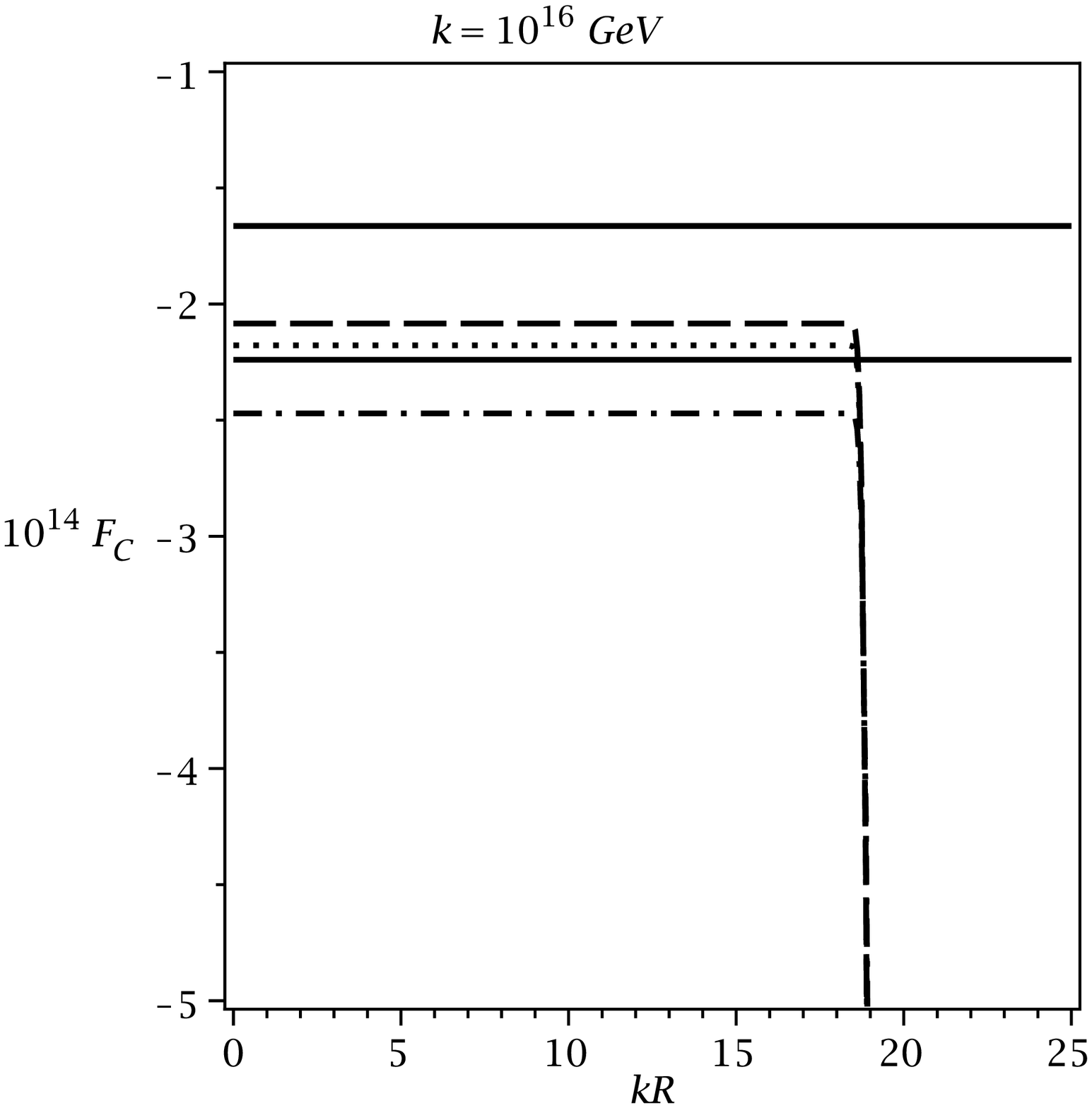}
\par\end{centering}

\vskip -0.2in 

\caption{The Casimir force as a function of the interbrane distance $kR$ in
the RSI model for plate separation $L=0.5\mu$m. a) $k=10^{19}$ GeV
and b) $k=10^{16}$ GeV. The dash, dot and dash-dot curves are for
$l_{nc}=0.01,\:0.05$ and $0.1\,\mu$m, respectively. }

\end{figure}

Before ending this analysis on the nature of the Casimir force in
the NC RSI brane scenario, let us comment about the recent finding
of the author of ref.\cite{Cheng}. Discarding the contribution of
the force without the braneworld scenario, the author claims that
the Casimir force in the RSI model becomes repulsive when $\mu<0.156$.
We have performed the calculation starting from Eq.($\ref{force}$)
with $\lambda=0$, and we have observed no deviation from the attractive
character of the Casimir force. It is only the presence of a NC cutoff
which reveals the repulsive character of the Casimir force in the
small plate separation regime. Finally, the numerical analysis of
the Casimir force in the NC RSII model shows no observable deviation
from the standard attractive Casimir force even with a repulsive NC
contribution. As an indication of this observation, we have taken
$L=0.5\:\mu$m and $k=10^{19}$ GeV and found that the correction
term to the standard Casimir force is of the order of $10^{-40}.$

%%%%%%%%%%%%%%%%%%%%%%%%%%%%%%%%%%%%%%%%%%%%%%%%%%%%%%%%

\section{Conclusion}

In this paper, we have have analyzed the effects of spacetime noncommutativity
on the Casimir force in 5-dimensional Randall and Sundrum (RS) braneworld
models. We derived to leading order in the NC length, which corresponds
to the case where the NC length is smaller than the interbrane distance,
 the expressions of Casimir energy and Casimir force for bulk electromagnetic
field confined between parallel plates in the RSI and RSII models.
Using accurate experimental measurements of the Casimir force for
parallel plates geometry to set bounds on the NC cutoff, we found
that for a cutoff of the order of $\Lambda_{nc}>10^{3}$ TeV, the
NC part of the Casimir force contributes significantly and that the
force develops a repulsive part for small plate separation, whose
location depends on the value of the NC cutoff. However, we found
that for $\Lambda_{nc}\lesssim10^{3}$ TeV, the Casimir force is always
attractive. Finally , analyzing the variation of the Casimir force
as a function of the interbrane distance $kR$ for fixed plate separation
and different values of the NC cutoff, we found that $kR$ $\lesssim20.5$
and $kR\lesssim18.4$ for $k=10^{19}$ GeV and $k=10^{16}$ GeV, respectively,
confirming the previous finding of \cite{Turan}.

We also performed the calculation of the Casimir force in the NC RSII
brane model, in which the 3-brane at the boundary $y=\pi R$ is at
infinity. We obtained an analytical expression for the Casimir force
between the plates, which shows that the force contains a repulsive
NC contribution but remains attractive, and that the deviation from
the experimentally measured force is too small to be significant.
This supports previous findings that RSII brane scenario has no low
energy measurable consequences. 

%%%%%%%%%%%%%%%%%%%%%%%%%%%%%%%%%%%%%%%%%%%%%%%%%%%%%%%%

\section*{Acknowledgment}

One of the authors (K. N) thanks the Algerian Ministry of Scientific
Research and High Education for financial support and the Frankfurt
Institute for Advanced Studies (FIAS), specially Professor Walter
Greiner for warm hospitality.

%\bibliography{}

%%%%%%%%%%%%%INTRO%%%%%%%%%%%%%%%%%%%%%%%%%%%

\end{document}